\documentclass[
journal=jctcce,
manuscript=article,
layout=onecolumn
]{achemso}
\usepackage[T1]{fontenc}
\usepackage{tocloft}
\usepackage{graphicx}  
\usepackage{dcolumn}   
\usepackage{bm}        
\usepackage{amssymb}   
\usepackage{amsmath}
\usepackage{mathtools}
\usepackage{cancel}
\usepackage{physics}
\usepackage{float}
\usepackage{afterpage}
\usepackage[normalem]{ulem}
\usepackage{hyperref}
\usepackage{placeins}
\usepackage{chemformula}
\hypersetup{
     colorlinks=true,
     linkcolor=blue,
     filecolor=blue,
     citecolor = black,      
     urlcolor=blue,
     }

\usepackage{xfrac}

\usepackage{orcidlink}
\usepackage{xcolor}
\definecolor{amber}{rgb}{1,0.49,0}

\newcommand{\editorr}[2]{%
  \expandafter\newcommand\csname #1note\endcsname[1]{%
    \textcolor{#2}{(\textbf{#1:} ##1)}}%
  \expandafter\newcommand\csname #1\endcsname[1]{%
    \textcolor{#2}{##1}}%
  \expandafter\newcommand\csname #1cancel\endcsname[1]{%
    \textcolor{#2}{\sout{##1}}}%
  \expandafter\newcommand\csname #1change\endcsname[2]{%
    \textcolor{#2}{\sout{##1} ##2}}%
  \newenvironment{#1text}{\color{#2}}{\color{black}}
}

\editorr{SB}{amber}
\editorr{ED}{blue}
\editorr{CAVEAT}{red}
\editorr{resub}{cyan}

\renewcommand{\[}{\begin{equation}}
	\renewcommand{\]}{\end{equation}}
\renewcommand{\(}{\begin{equation*}}
	\renewcommand{\)}{\end{equation*}}

\def\mm#1{\left\langle#1\right\rangle}

\DeclarePairedDelimiterX\set[1]\lbrace\rbrace{#1}

\newcommand{\kk}{{\bm k }}
\newcommand{\rr}{{\bm r }}

\providecommand{\keywords}[1]
{
	\small	
	\textbf{\textit{Keywords---}} #1
}

\title{Seebeck coefficient of liquid water \\ from equilibrium molecular dynamics}
	
\author{Enrico Drigo\, \orcidlink{0000-0002-1797-2987}}
\affiliation[sissa]{SISSA – Scuola Internazionale Superiore di Studi Avanzati, Via Bonomea 265, 34136 Trieste, Italy}

\email{endrigo@sissa.it}

\author{Stefano Baroni\,\orcidlink{0000-0002-3508-6663}}
\affiliation[sissa]{SISSA – Scuola Internazionale Superiore di Studi Avanzati, Via Bonomea 265, 34136 Trieste, Italy}
\alsoaffiliation[cnr]{CNR-IOM DEMOCRITOS, SISSA, Via Bonomea 265, 34136 Trieste, Italy}

\keywords{polarization $|$ thermal gradient $|$ molecular simulation $|$ equilibrium thermodynamics} 

\begin{document}
\begin{abstract}
 The application of a temperature gradient to an extended system generates an electromotive force that induces an electric current in conductors and a macroscopic polarization in insulators. The ratio of the electromotive force to the temperature difference, usually referred to as the Seebeck coefficient, is often computed using non-equilibrium techniques, such as non-equilibrium molecular dynamics (NEMD). In this paper we argue that thermo-polarization effects in insulating fluids can be conveniently treated by standard equilibrium thermodynamics and devise a protocol---based on a combination of equilibrium molecular dynamics and Bayesian inference methods---that allows one to compute the Seebeck coefficient in these systems along with a rigorous estimate of the resulting statistical accuracy. The application of our methodology to liquid SPC/E water results in good agreement with previous studies---based on more elaborate NEMD simulations---and in a more reliable estimate of the statistical accuracy of the results.
\end{abstract}
%

	
\section*{Introduction}
The coupling between temperature and electric fields in extended systems characterizes non-equilibrium phenomena of great relevance in physics and materials science. Together with the Peltier and Thompson effects, the Seebeck effect is one of the most common thermo-electric phenomena \cite{Snyder2008}. The Seebeck coefficient describes the electromotive force induced by a temperature gradient in open-circuit conditions, \emph{i.e.} when no electric current flows. While in metals and electrolytes this electromotive force generates an electric current when the circuit is closed, in liquid insulators it is generated by a macroscopic electric polarization, which, in the case of liquid water, has been computed by several authors in the past decade or so, using non-equilibrium (NE) molecular dynamics (MD) simulations \cite{Bresme2008,bresme2013,bresme2016,armstrong2015,wirnsberger2016,wirnsberger2017}.
In this approach, a temperature gradient is mimicked by partitioning the simulation cell into two regions, whose edges act as heat sources and sinks. The thermo-polarization coefficient is then estimated from the magnitude of the electric field developed in the middle of the two regions, once the steady state has been attained. A full description of the response of the system to a temperature gradient entails therefore a proper account of the heat current generated by it.
	
In linear response theory (LRT), the response of the macroscopic charge and heat currents, $\bm{J}_e$ and $\bm{J}_q$, to an electric field, $\bm{E}$, and a temperature gradient, $\grad T$, is expressed through the Onsager constitutive equations \cite{onsagerI1931,onsagerII1931}:

\begin{equation}
    \begin{aligned}
	    \bm{J}_e &= \sigma\bm{E} - K_{12} \frac{\grad T}{T}, \\
	    \bm{J}_q &= K_{12}e\bm{E} - L_{qq}\frac{\grad T}{T},
    \end{aligned}
    \label{eq:Onsager}
\end{equation}
where $\{\sigma, K_{12}, L_{qq}\}$ are the Onsager coefficients, as defined, \emph{e.g.}, in Refs. \cite{kubo1957a,kubo1957b, martin1967}.
By definition, the Seebeck coefficient, $S$, is the ratio between the electric field and the temperature gradient, as measured when no electric current flows. By imposing this condition in Eqs. \ref{eq:Onsager}, $S$ can be written as \cite{martin1967, callen1948}:

\begin{equation}
	S\doteq\left.\frac{{E}}{\grad T }\right\vert_{\bm{J} =0}=\frac{K_{12}}{\sigma T}, \label{seebeckonsager}
\end{equation}
where $E$ and $\grad T$ are any Cartesian components of the electric field and temperature gradient, respectively, and space isotropy is assumed throughout. Heat and charge currents are odd with respect to time-reversal symmetry, whereas the thermodynamic forces (electric field and temperature gradient) that determine them are even, thus apparently violating the time-reversal invariance of the system and giving rise to dissipation and the macroscopic irreversibility of microscopically reversible systems. Mathematically, this outward conundrum is due to the non-commutativity of the low-frequency and long-wavelength limits of the response functions relating currents to forces \cite{kubo1957a,kubo1957b, forster}. As a result, the Onsager coefficients are usually computed from LRT as integrals of time correlation functions, \emph{i.e.} as low-frequency limits of response functions when the wavelength of the perturbation goes to infinity (in this order). Since electric fields and temperature gradients have the same parity under time reversal, it is to be expected that the order of the limits does not matter in this case, and that their ratio in \eqref{seebeckonsager} can be expressed as the long-wavelength limit of a suitable \emph{static} response function. In insulators, charge transport is forbidden and thus the associated transport coefficients vanish. In this regime, the ratio in \eqref{seebeckonsager} is numerically ill-conditioned making the computation of the Seebeck coefficient via static response functions particularly convenient. Physically, the ability of equilibrium thermodynamics to account for the polarization response to a temperature gradient---in spite of the concurrent off-equilibrium heat-flux response to the thermal perturbation---depends on the adiabatic decoupling between dielectric relaxation, which occurs on a molecular time scale, and thermal relaxation, that occurs over hydrodynamic times.

In this paper we will follow this path and show that the Seebeck coefficient can be expressed as the long-wavelength limit of the heat-density/charge-density equal-time correlation function. In order to evaluate this limit from equilibrium MD without resorting to finite-size extrapolations, we devise a Bayesian technique that also provides for a rigorous estimate of the statistical accuracy of the procedure. Our methodology is demonstrated on liquid water in a wide range of temperatures and pressures. The results are in fair agreement with previous theoretical estimates based on NEMD, whose efficiency and statistical accuracy are not as easy to assess. 

	
\section*{Results and discussion}
\subsection*{Theory}
In the absence of external charges, the electric field is entirely due to the polarization  of the system, $\bm P$: $\bm E=-4\pi \bm P$. The Seebeck coefficient reads therefore:

\begin{equation}
    S = -4\pi \frac{P}{\grad T} \label{eq:Seebeck-1}
\end{equation}
and is thus proportional to the linear response of the electric polarization to a temperature gradient.

In order to apply Hamiltonian perturbation theory to the linear response of an extended system to a thermal disturbance, such as a temperature gradient, it is expedient to describe the latter in terms of an equivalent mechanical perturbation, $\widehat{\mathcal{V}}$, which, to lowest order, induces the same unbalance in the energy-density distribution \cite{Luttinger1964,StefanoHandbook2018}:

\begin{equation}
    \begin{aligned}
	   \widehat{\mathcal{V}} &= -\frac{1}{T}\int  \hat{q}(\rr) T'(\rr) d\rr, \\
        &= -\frac{V}{T}\sum_\kk \widehat{\widetilde q}(\kk) \widetilde T'(-\kk),
    \end{aligned}
    \label{eq:V-hat}
\end{equation}
where  $T$ is the average temperature of the system, $V$ the system's volume, $T'(\rr) = T(\rr)-T$ the departure of the local temperature from its space average, and $\hat{q}(\rr) = \hat{e}(\rr) - \sum_i h_i\hat{n}_i(\rr)$ is the heat density, $\hat{e}(\rr)$ being the energy density, $\hat{n}_i(\rr)$ the number density of the $i$-th molecular species, and $h_i$ the corresponding partial enthalpy. In Eqs. \ref{eq:V-hat} a caret indicates an implicit dependence on phase-space variables, $\Gamma$, as in $\widehat{\mathcal{V}} = \mathcal{V}(\Gamma)$, and a tilde a Fourier transform, defined as $\widetilde F(\kk)= \frac{1}{V}\int_VF(\rr)e^{-i\bm k\cdot \rr}d\rr$.

A uniform temperature gradient perturbation is not compatible with periodic boundary conditions (PBCs), which are commonly used in molecular simulations, and it must be intended as the long-wavelength limit of a periodic perturbation of wave-vector $\kk$. Because of Gauss' law, the Fourier transform of the polarization, $\widetilde{\bm P}(\kk)$, can be put into the form: $\widetilde {\bm P}(\kk)=i\kk \widetilde\varrho(\kk)/k^2$, $\widetilde\varrho(\kk)$ being the Fourier transform of the charge-density distribution, $\varrho(\rr)$. The Fourier transform of the temperature gradient reads: $\frac{1}{V}\int_V\grad T(\rr) e^{-i\kk\cdot\rr}d\rr = i\kk \widetilde T(\kk). $ The Seebeck coefficient,  
\eqref{eq:Seebeck-1}, reads therefore:

\begin{align}
    S &=\lim_{\kk\to0} S(\kk),\text{ where} \label{eq:S-limit} \\ 
    S(\kk) &=- \frac{4\pi}{k^2}
    \widetilde\chi_{q\varrho}(\kk), \label{eq:S(k)}
\end{align}
and $\widetilde\chi_{q\varrho}(\kk)$ is the Fourier transform of the charge-temperature susceptibility. According to standard LRT, $\widetilde\chi_{q\varrho}(\kk)$ can be expressed as an equal-time correlation function between the charge and heat densities:

\begin{align}
    \begin{aligned}
        \widetilde\chi_{\varrho q}(\kk) &\doteq \frac{\partial \widetilde\varrho(\kk)}{\partial \widetilde T(\kk)} \\
        & =\frac{V}{k_BT^2} \left \langle \widehat{\widetilde {q}}(\kk)\widehat{\widetilde\varrho}(-\kk)\right \rangle_0, 
    \end{aligned}
    \label{eq:chi_rhoq}
\end{align}
where $k_B$ is the Boltzmann's constant and $\mm{\,}_0$ is an equilibrium average over the initial conditions of a molecular trajectory. 
The heat density, $q(\rr)$, entering \eqref{eq:chi_rhoq} is intrinsically ill-defined, as it is affected by a so-called \emph{gauge freedom}, deriving from the insensitivity of the total energy---and therefore of all the macroscopic thermal properties---on the addition of the divergence of a bounded vector field to the energy density: $e(\rr)\to e(\rr)+ \grad\cdot \bm p(\rr)$ \cite{Marcolongo2016,Ercole2017,Grasselli2021}. Though the $\kk= 0$ value of  \eqref{eq:chi_rhoq} is insensitive to $\bm p$, the limit in Eqs. (\ref{eq:S-limit}--\ref{eq:S(k)}) may depend on it. The dependence of correlation functions on the specific definition of the energy density at finite wavelength is to be expected, and it is in fact at the very root of energy-gauge freedom. While it is thus not surprising that higher derivatives of the correlation functions at $\kk=0$ may depend on the energy gauge, the deeper meaning of such dependence in the specific case of the Seebeck coefficient as defined by Eqs. (\ref{eq:S-limit}--\ref{eq:S(k)}) probably deserves further investigation. For the time being, suffice it to say that full gauge invariance is restored if one restricts the choice of the $\bm p$ gauge vector filed to the gradient of a scalar field.

\subsection*{Simulations for liquid water}
We have performed several equilibrium MD simulations of  $512$ rigid water molecules in a cubic box, using the SPC/E force field \cite{spce} with PBCs at different pressure and temperature conditions. In order to extrapolate the value of the limit in \eqref{eq:S-limit}  we fitted the $S(\kk)$ to a polynomial whose order and coefficients were determined via a Bayesian linear regression algorithm \cite{bishop2006}. All the details of the simulation and of the Bayesian data analysis procedure are discussed in the \emph{Methods} section.

\begin{figure}[t]
    \includegraphics[width=\linewidth]{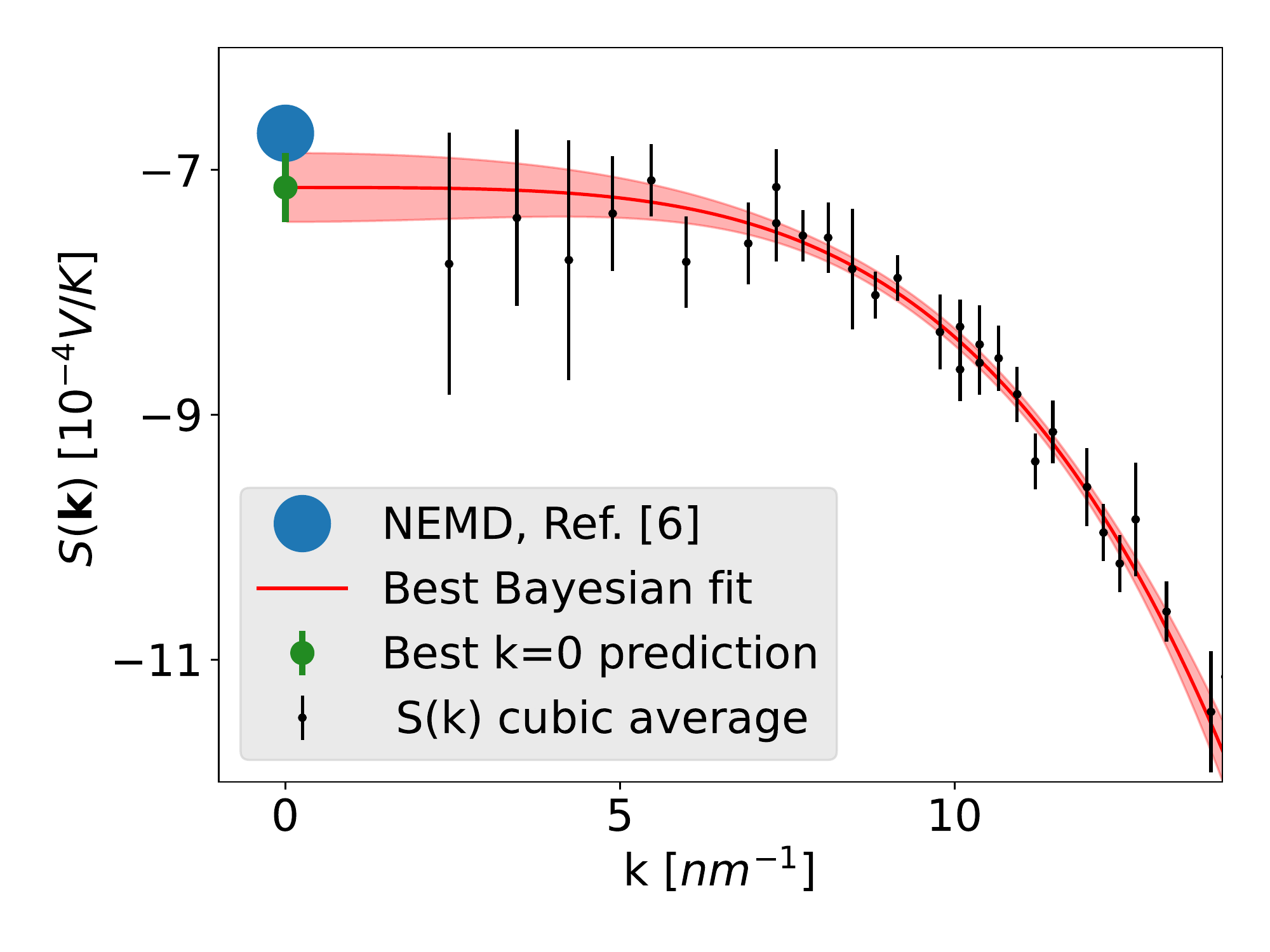}
    \caption{
    Wave-vector dependence of the Seebeck coefficient, $S(\kk)$,  \eqref{eq:S(k)}, of SPC/E water at $\mathrm T=400~K$ and zero pressure, as estimated from our MD simulations. The reported data are averages over different wave-vectors of the same magnitude, compatible with our PBCs. The red line is our Bayesian polynomial fit, with the shaded area indicating the estimated uncertainty. At $\kk=0$ we also report our extrapolation and the value obtained by Wirnsberger et al. \cite{wirnsberger2016} via NEMD.
    } \label{spcefit}
\end{figure}

	
  
In Fig. \ref{spcefit} we display the wave-vector dependence of the Seebeck coefficient as computed for our water model at $400$ K and zero pressure, along with our Bayesian polynomial fit that allows us to estimate the $\kk\to 0$ extrapolation, also reported. Our estimate of the Seebeck coefficient of SPC/E water at room pressure and $400$ K is $-0.71 \pm 0.03$  mV/K leading to an electric field of magnitude $\approx 10^7$ V/m for thermal gradients of $\approx 5$ K/\AA,\, in agreement with the literature \cite{wirnsberger2016,wirnsberger2017,Bresme2008,armstrong2015,bresme2013,bresme2016}.

\begin{figure}[t!]
    \includegraphics[width=\linewidth]{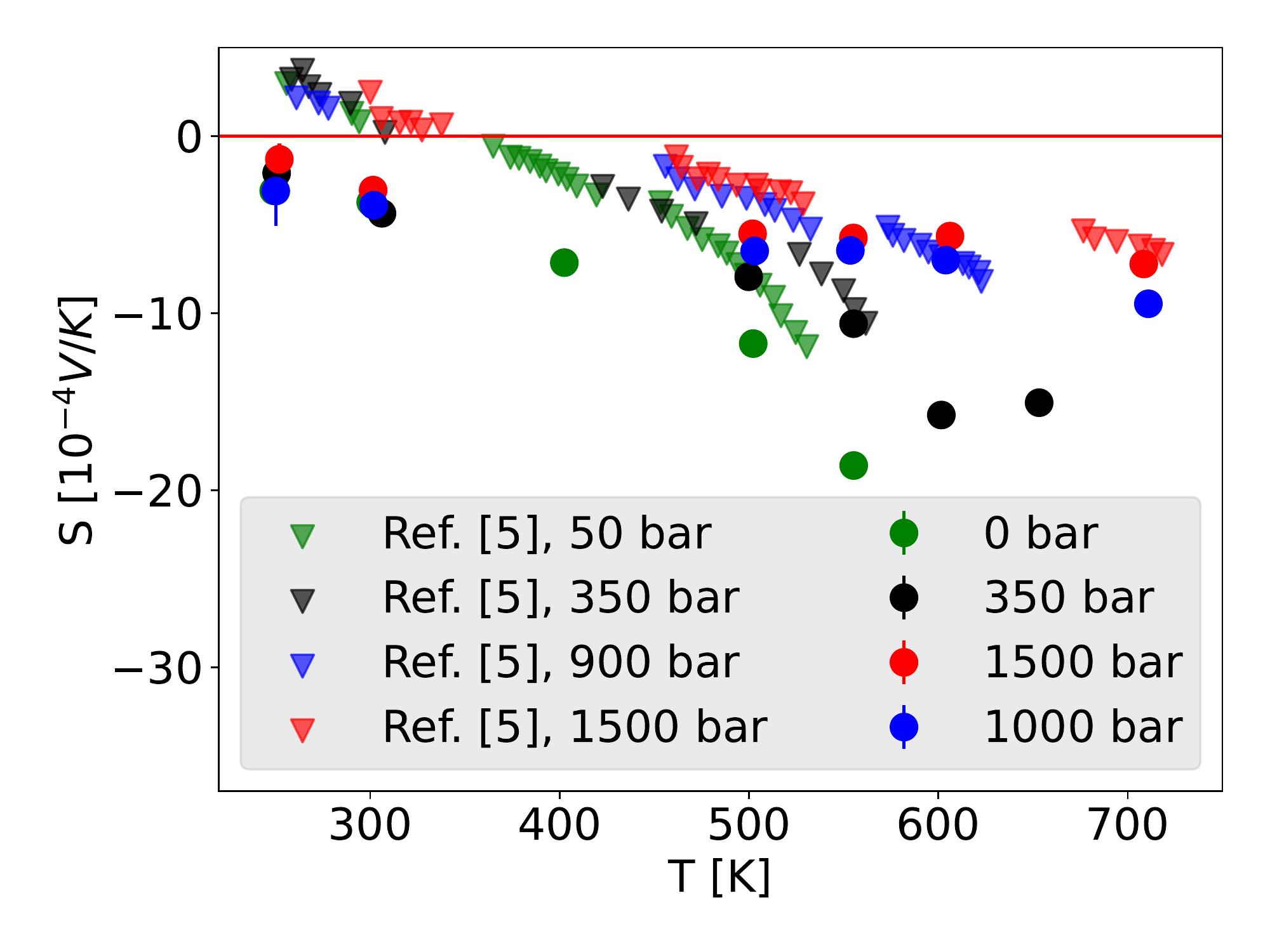}
    \caption{Seebeck coefficient, $S$, \eqref{eq:S-limit}, as estimated at various temperature and pressure conditions, using the Bayesian linear regression method described in the text. The results are compared with NEMD simulations of SPC/E water by Armstrong et al. \cite{armstrong2015}.} \label{spcephasediagrambresme}
\end{figure}

In Fig. \ref{spcephasediagrambresme} we report the results of our simulations, covering a broad range of pressures and temperatures, and compare them with the results of Armstrong and Bresme \cite{armstrong2015}. We observe that the pressure dependence of $S$ is stronger as the temperature increase, and its temperature dependence is weaker as the applied pressure increases. While these qualitative trends are in agreement with previous results obtained from NEMD \cite{armstrong2015}, we do not find any evidence of the temperature inversion of the thermal polarization in water observed in Ref. \cite{armstrong2015}.

\section*{Methods}
\subsection*{Molecular dynamics}
We represented bulk water with a sample of 512 molecules with PBCs over a cubic simulation cell, using the SPC/E rigid model \cite{spce}: \ch{O}-\ch{H} bond lengths and \ch{H}-\ch{O}-\ch{H} bond angles are kept fixed and inter-atomic forces are described by a combination of Lennard-Jones plus Coulomb potentials, whose parameters are fitted to experimental data.
The resulting static and dynamic properties, such as the radial distribution function and the diffusivity, are in great agreement with the experiments. Molecular trajectories were generated using the velocity-Verlet algorithm \cite{allen2017computer} with a time step of $0.25$ fs using the parallel code \texttt{LAMMPS} \cite{Lammps, lammps1}. The long-range Coulomb interaction was treated via the Ewald summation method with a real-space cut-off radius of 11 Å. All data are harvested from 500-ps long molecular NVE trajectories, after careful equilibration performed in a number of different ensembles. The $\kk$-dependent Seebeck coefficient, as well as other correlation functions computed for benchmarking purposes, were evaluated on a uniform grid of reciprocal-lattice vectors compatible with the PBCs in use.


\begin{figure}
    \includegraphics[width=\linewidth]{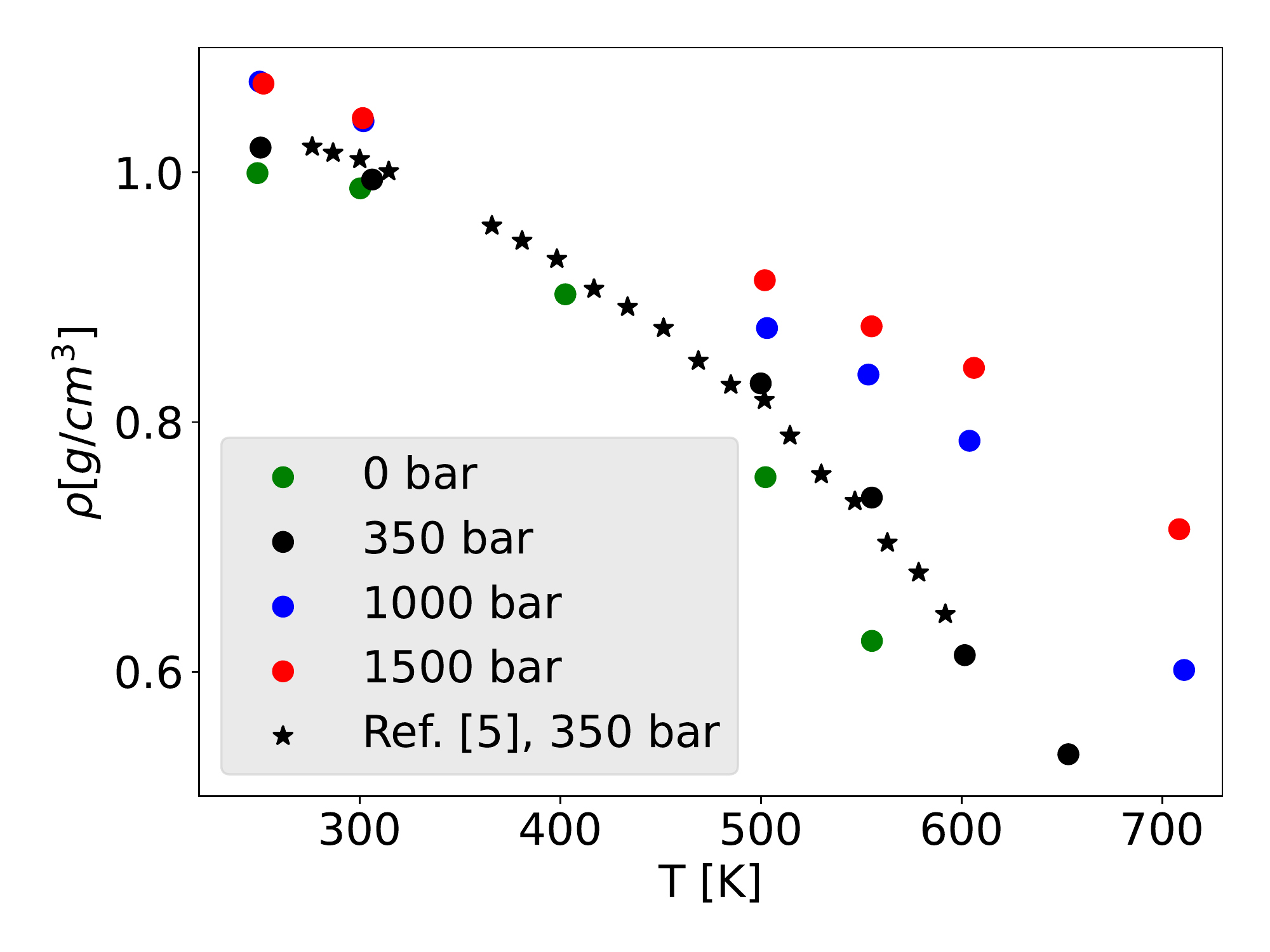}
    \caption{Temperature dependence of the density of SPC/E water evaluated at different pressures and compared with the results of Armstrong et al. \cite{armstrong2015} in the specific case of $p=350~\mathrm{bar}$.
    } \label{spcerhobresme} 
\end{figure}

The temperature dependence of the density for SPC/E water was estimated at different pressures and compared with the results obtained in Ref. \cite{armstrong2015} at $350$ bar, resulting in a substantial agreement (see Fig. \ref{spcerhobresme}).



\subsection*{Bayesian linear regression}\nobreak

In order to evaluate the limit in \eqref{eq:S-limit}, we fit the estimated dependence of the Seebeck coefficient upon wavenumber, $S(\kk)$, \eqref{eq:S(k)}, to a low-order polynomial. Isotropy dictates that $S(\kk)$ is actually a function of the squared modulus of its argument, $k^2$.

\begin{equation} \label{eq:polynomial_fit}
    S(\kk) \approx w_0+w_1 k^2+w_2 k^4\dots+w_{M}k^{2M}
\end{equation}
In practical simulations, rotational symmetry is reduced from spherical to cubic, due to the use of PBCs. As a consequence, slightly different values of $S(\kk)$ may correspond to different wavevectors of equal magnitude. In our analysis of the data generated by MD simulations this artifact is corrected by averaging the values of $S(\kk)$ over all the wavevectors of equal magnitude. We have checked that explicitly accounting for cubic symmetry would not change the final estimate of the Seebeck coefficient, while slightly increasing the resulting uncertainty. In order to estimate the coefficients of the polynomial fit in  \eqref{eq:polynomial_fit}, $\bm{w}=\{w_0\cdots w_M\}$, we resort to a Bayesian inference method, starting from the likelihood function, $\mathcal{L}(\bm{w})=\sum_i \frac{\left(S_i - \bm{w}\cdot\bm{\Phi}_i\right)^2}{2\sigma_i^2}$, where $S_i$ is our MD estimate of the Seebeck coefficient at wavevector $\kk_i$, $\bm{\Phi}_i$ is the basis set  consisting of  monomials of even degree, $\bm{\Phi}_i=\bm \Phi(\kk_i)\equiv \{1,k_i^2\dots k_i^{2M}\}$, evaluated at the wavevectors being sampled, and $\sigma_i$ is the standard deviation of $S_i$, as estimated via block analysis. The probability that the data-set, $\mathcal{D}\equiv\{S_i\}$, is generated from the function $\bm{w}\cdot \bm{\Phi}_i$
is proportional to the exponential of the negative of the likelihood, $p\left(\mathcal{D}\vert \bm{w}\right)\propto \exp\left [- \mathcal{L}(\bm{w}) \right ]$. We determine the posterior distribution, $p\left( \bm{w}\vert\mathcal{D}\right)$, i.e. the probability that $\bm{w}\cdot \bm{\Phi}(\kk)$ is the correct function from which  $\mathcal{D}$ was sampled, by leveraging the Bayes theorem: $p\left(\bm{w}\vert \mathcal{D}\right)p(\mathcal{D})=p\left(\mathcal{D}\vert \bm{w} \right)p(\bm{w})$. Assuming that the prior distribution is normal, $p(\bm w)=\left(\frac{\alpha}{2\pi}\right)^{\frac{M}{2}}\exp\left [-\alpha\norm{\bm{w}}^2 \right ]$, is equivalent to introducing a  regularization term, $\alpha\norm{\bm{w}}^2$, in ordinary linear regression, so as to prevent over-fitting. The prior distribution of the parameters depends implicitly on their number, $M+1$, and on the hyper-parameter $\alpha$. By leveraging again Bayes theorem, the optimal values of $M$ and $\alpha$ are determined as those that maximize the probability of their occurrence, conditionally to actual observation of the data set:

\begin{equation}
    p(M,\alpha\vert \mathcal D)\propto p(\mathcal D|M,\alpha) p(M,\alpha).
\end{equation}
In order to proceed further, we assume a flat distribution for the $(M,\alpha)$ prior, $P(M,\alpha)\approx \text{cnst}$, and express $p(\mathcal D|M,\alpha)$ as a marginal distribution:

\[p\left(\mathcal{D}\vert M, \alpha \right)=\int  p\left(\mathcal{D}\vert \bm{w}\right)  
	p\left(\bm{w}\vert M, \alpha\right) d\bm{w}.
 \] 
 %
The actual procedure that we followed to determine $M$, $\alpha$, and $\bm w$, and therefore $S=w_0$, is described in  Ref. \cite{bishop2006}. 

\begin{figure}[t]
    \includegraphics[width=\linewidth]{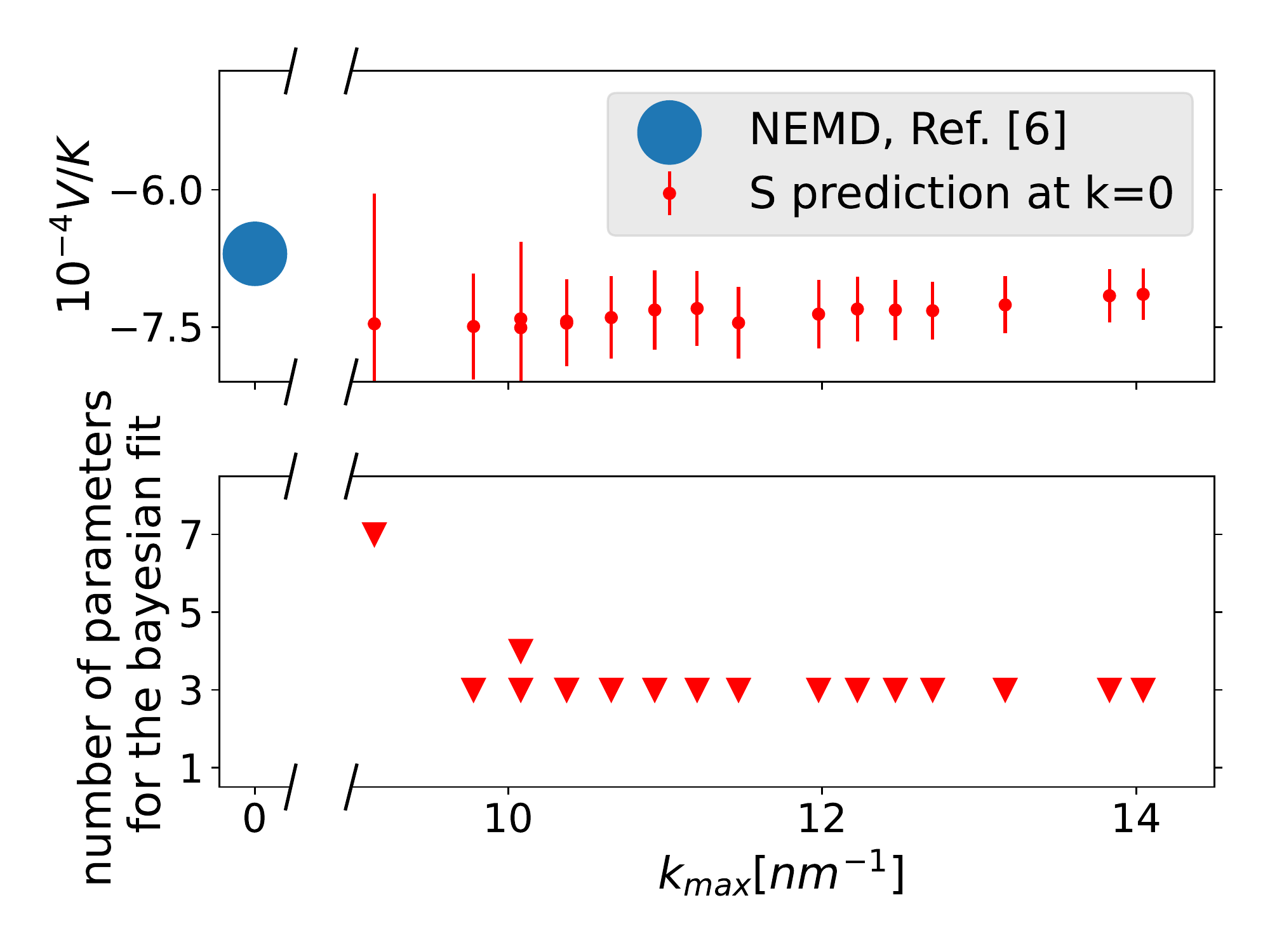}
    \caption{
        Upper panel: Bayesian regression prediction of the  $\kk=0$ value of the Seebeck coefficient in SPC/E water at $400$ K and $0$ bar, as a function of the maximum $\kk$-vector, $k_{max}$, included in the data set.
        The 
        NEMD result from Ref. \cite{wirnsberger2016} is also reported. Lower panel: Number of parameters of the Bayesian regression fit as a function of $k_{max}$
        } \label{spcepred} 
\end{figure}

In Fig. \ref{spcepred} we report the value of the ($\kk=0$) Seebeck coefficient estimated from our MD simulation, as a function of the maximum wavevector included in the dataset for the Bayesian-inference analysis.
Increasing the number of $\kk$-points in the data-set does not change the prediction, meaning that the procedure is stable and consistent. The Bayesian extrapolation is in accordance with previous NEMD calculations of Ref. \cite{wirnsberger2016}, whose results are also reported in the figure.

\begin{figure}[h!]
    \includegraphics[width=0.6\linewidth]{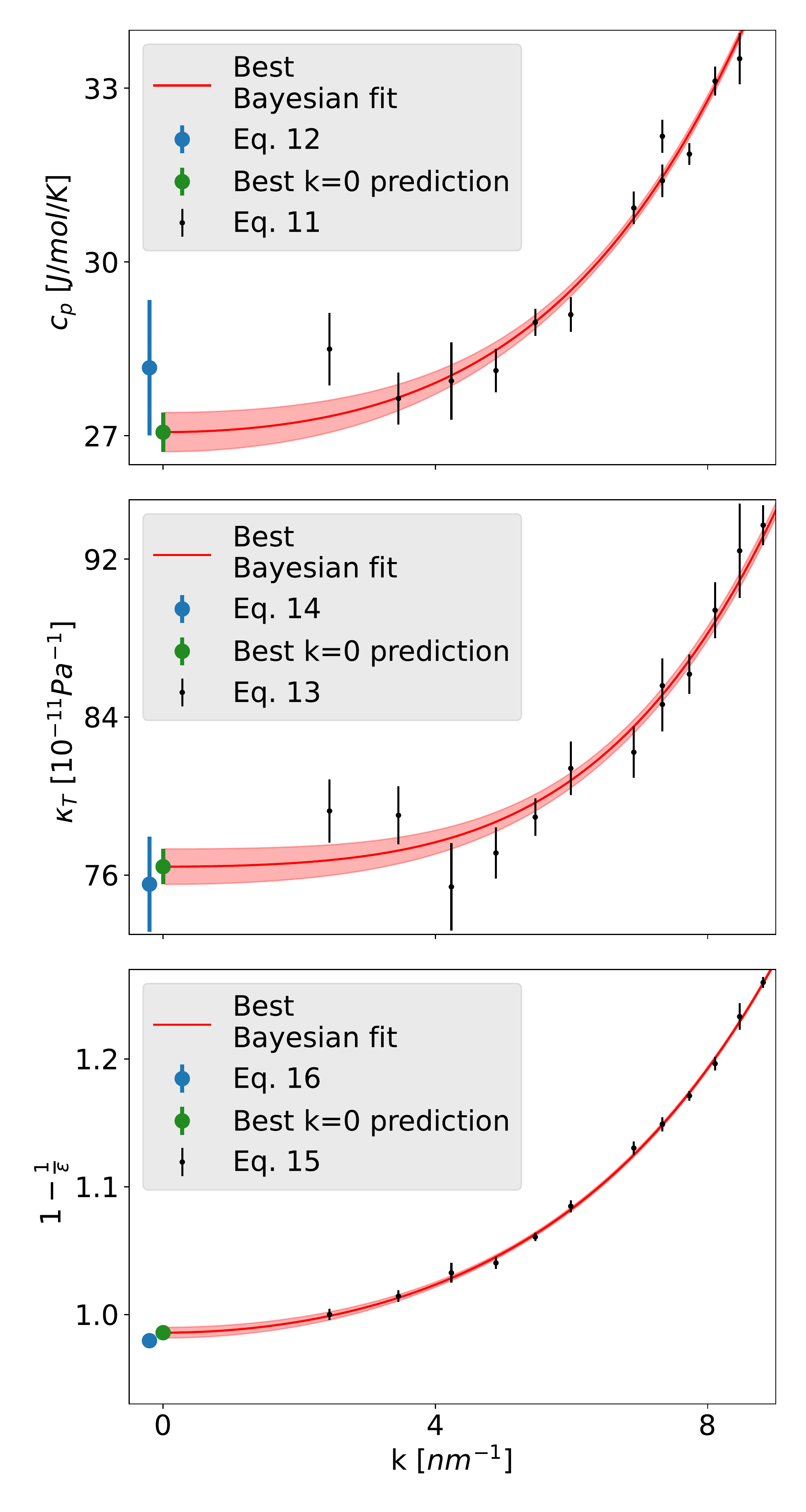}
    \caption{
    Bayesian-inference estimates of the isobaric specific heat, $c_p$, isothermal compressibility, $\kappa_T$, and dielectric constant, $\epsilon$, of liquid SPC/E water at 400 K and 0 bar, Eqs. (\ref{eq:cp_k}-\ref{eq:eps_k=0}).
    } \label{cpktpol} 
\end{figure}

In order to validate our Bayesian inference procedure, we have applied it to the estimate of the long-wavelength limit of other response functions, which could also be evaluated from fluctuations of lattice-periodic ($\kk=0$) observables, such as the constant-pressure specific heat, $c_p$, the isothermal compressibility, $\kappa_T$, or the dielectric constant, $\epsilon=1+4\pi\chi$, $\chi=\frac{\partial P}{\partial E}$ being the macroscopic polarizability of the system. Standard fluctuation theory gives \cite{allen2017computer, forster}:

\begin{alignat}{4}
    &c_p &&= \frac{1}{N}\left ( \frac{\partial H}{\partial T} \right )_{p} &&=\frac{V}{nk_B T^2}\lim_{\kk\to 0} \langle \widehat{\widetilde {q}}(\kk)\widehat{\widetilde {q}}(-\kk)\rangle_{\scriptscriptstyle NVE}, \label{eq:cp_k}  \\ 
    &&&&& =\frac{1}{Nk_B T^2}\left \langle \Delta H^2 \right\rangle_{\scriptscriptstyle NPT}, \label{eq:cp_k=0} 
\end{alignat}
\vspace{-\belowdisplayskip}\vspace{-\abovedisplayskip}

\begin{alignat}{4}
    &\kappa_T&&=-\frac{1}{V}\left ( \frac{\partial V}{\partial p}\right )_T &&=\frac{V}{n^2k_B T}\lim_{\kk\to 0} \langle \widehat{\widetilde {n}}(\kk)\widehat{\widetilde {n}}(-\kk)\rangle_{\scriptscriptstyle NVE}, \label{eq:kappaT_k}\\
    &&&&& = \frac{1}{Vk_B T}\left \langle \Delta V^2 \right\rangle_{\scriptscriptstyle NPT} \label{eq:kappaT_k=0}
\end{alignat}
\vspace{-\belowdisplayskip}\vspace{-\abovedisplayskip}

\begin{alignat}{4}
    &1-\frac{1}{\epsilon} &&= \frac{\partial P}{\partial D} &&=\frac{4\pi V}{ k_B T}\lim_{\kk\to 0} \frac{\langle \widehat{\widetilde {\varrho}}(\kk)\widehat{\widetilde {\varrho}}(-\kk)\rangle_{\scriptscriptstyle NVE,\bm D}}{k^2} \label{eq:eps_k} \\
    &&&&& = 1- \frac{k_B T}{4\pi V\langle P^2 \rangle_{\scriptscriptstyle NVE,\bm E}} \label{eq:eps_k=0}
\end{alignat}
where $N$ is the number of molecules and $n=N/V$ 
the molecular density, ``$NVE$'' and ``$NPT$'' indicate the micro-canonic and iso-baric/iso-thermal ensembles, respectively, and ``$NVE, \bm E$'' and ``$NVE, \bm D$'' indicate the microcanonic ensembles where the (total) electric field and electric induction (i.e. applied field) are kept constant, respectively. Notice that molecular simulations performed with PBC tacitly assume $\bm D(\kk\ne 0)=0$, i.e. no spatially varying applied field, and $\bm E(\kk= 0)=0$, i.e. it is the macroscopic component of the \emph{total}, rather than \emph{applied}, electric field that is assumed to vanish. This last condition is a consequence of the use of PBCs, which, in the Coulomb gauge usually adopted in molecular simulations, are incompatible with a finite value of the total macroscopic electric field. Therefore, the fluctuation of the squared magnitude of the Fourier transform of the electric polarization, $\langle|\bm P(\kk)^2|\rangle$ makes a jump when passing from $\kk_{min}$ to $\kk=0$. Note that \eqref{eq:eps_k=0} can only be implemented as such when a well defined value for microscopic dipole can be associated to each molecule, so that the macroscopic ($\kk=0$) polarization of the system has a well defined meaning. When this is not possible, e.g. in an \emph{ab initio} quantum mechanical setting or, more generally, when molecular bonds are allowed to break and form, the total dipole of a system in PBC cannot be rigorously defined, and only \eqref{eq:eps_k} can be given a rigorous meaning and practical implementation.

In Fig. \ref{cpktpol} we illustrate our Bayesian extrapolation procedure to estimate the $\kk\to 0$ limits of the correlation functions delivering the constant-pressure specific heat, isothermal compressibility, and dielectric constants, Eqs. (\ref{eq:cp_k},\ref{eq:kappaT_k},\ref{eq:eps_k}), which results to be in very good agreement with the predictions of fluctuation theory at $\kk=0$, Eqs. (\ref{eq:cp_k=0},\ref{eq:kappaT_k=0},\ref{eq:eps_k=0}).


\section*{Conclusions}
On a more fundamental side, we believe that our work highlights a conceptual distinction between genuinely off-equilibrium phenomena, such as transport properties, and others that---while usually treated as such and indeed inextricably tangled with them---can be treated as (quasi-) equilibrium ones. Thermo-polarization is one such phenomenon in that---while thermal equilibrium cannot be achieved in the presence of a temperature gradient---the time scale of the dielectric relaxation responsible for the polarization of the medium is much faster than the hydrodynamic time scale characteristic of thermalization. It should thus not come as a surprise that the polarization response to a temperature gradient---while resulting from the same disturbance that also gives rise to heat transport---can be conveniently treated by equilibrium thermodynamic techniques.

On a more practical side, we have proposed an elementary Bayesian inference method that we believe will be broadly applicable in all those cases where the long-wavelength limit of any function---such as e.g. a response function or a derivative thereof---cannot be evaluated directly as the expectation value of a periodic observable or correlation function, but can only be extrapolated from the values that the function acquires at finite and discrete wavevectors, $\kk$. One notable such application to the evaluation of thermal transport coefficients from energy-density fluctuations---rather than from current fluctuations as usually done---will be presented elsewhere.

Finally, on the applicative side, we notice that though the thermo-polarization effects discussed in this paper are tiny, they may become relevant in the presence of large temperature gradients, particularly if complex polar molecules enhance the electric susceptibility of the fluid. Such a situation has been recently claimed to occur in mitochondria where a temperature drop of more than 10 $^\circ\mathrm C$ was allegedly observed across the intermembrane space whose extent is of the order of 10 nm \cite{chretien2018}. Although this claim has been tempered by other authors \cite{macherel2021, di2022}, cellular metabolism---whereby a large power is developed in a confined complex molecular environment---may be the source of strong thermal gradients and important thermoelectric effects. Our simulations predict that an electromotive force of 2--6 mV is generated by a 4--12 $^\circ\mathrm C$ temperature drop across a 12-nm slab of pristine water at room temperature, representative of the alleged intermembrane temperature drop. The presence of complex polar molecules in the intermembrane medium may make these effects significant if the observed huge temperature drop is confirmed. Too little is known about the actual temperature distribution within a living cell and the mechanisms possibly giving rise to it to draw any conclusions, but whatever these mechanisms may turn out to be, heat transport and, maybe, thermoelectricity will play a relevant role in them, possibly opening yet unexplored avenues for molecular simulations in the life sciences.

\section*{Data Availability}
All study data are included in the article.

\begin{acknowledgement}
The authors are grateful to Riccardo Bertossa, Alfredo Fiorentino, Federico Grasselli, Maria Grazia Izzo, and Paolo Pegolo for many insightful discussions and valuable  suggestions. We also thank  Michele Vendruscolo and Daan Frenkel for a critical reading of our manuscript prior to publication. This work was partially supported by the European Commission through the \textsc{MaX} Centre of Excellence for supercomputing applications (grant number 101093374) and by the Italian MUR, through the PRIN project \emph{FERMAT} (grant number 2017KFY7XF) and the Italian National Centre for HPC, Big Data, and Quantum Computing (grant number CN00000013).
\end{acknowledgement}

\bibliography{biblio}
	
	
	

\end{document}